\documentstyle[prb,aps,epsf,multicol,epsfig]{revtex}
\begin{document}

\title{Pressure Dependence of the Irreversibility Line in 
Bi$_{2}$Sr$_{2}$CaCu$_{2}$O$_{8+\delta}$: 
Role of Anisotropy in Flux-Line Formation}

\author{M. P. Raphael$^*$, M. E. Reeves$^{\dagger,\ddagger }$, E. F. 
Skelton$^{\dagger,\ddagger}$, C. Kendziora$^{\ddagger}$}
\address{$^*$Catholic University of America, Washington, DC 20064}
\address{$^{\dagger}$The George Washington University, Washington, DC 20052}
\address{$^{\ddagger}$Naval Research Laboratory, Washington, DC 20375}
\maketitle
\begin{abstract}
One of the important problems of high-temperature superconductivity is to 
understand and ultimately to control
 fluxoid motion.  We present the results of a new technique for measuring the 
pressure dependence of 
the transition to superconductivity in a diamond 
anvil cell.  By measuring the third harmonic of the {\it ac } susceptibility, we 
determine the onset 
of irreversible flux motion.  This enables us to study the effects of pressure 
on flux motion.  The application of pressure
 changes interplanar 
spacing, and hence the interplanar 
coupling, without significantly disturbing the intraplanar superconductivity.  
Thus we are able to separate the 
effects of coupling from other properties that might affect the flux motion.
Our results directly show the relationship between lattice spacing, effective-
mass 
anisotropy, and the irreversibility line in 
Bi$_{2}$Sr$_{2}$CaCu$_{2}$O$_{8+\delta}$.  
Our results also demonstrate that an application of 2.5 GPa pressure causes 
a dramatic increase in interplanar coupling. 
\end{abstract}
\begin{multicols}{2}
\narrowtext
The strongly two dimensional nature of Bi$_{2}$Sr$_{2}$CaCu$_{2}$O$_{8+\delta}$  
makes it a particularly
apt system in which to study flux dynamics.  Theoretically, it has been 
recognized that 
the large anisotropy parameter and the temperature dependence of the Josephson 
coupling between planes
 should conspire to cause a crossover from vortex lines to 
pancakes.\cite{nelson89,fisher289,blatterrmp}  That is,
 the vortices change their topology from 3-D tubes to 2-D disks.  
Experimental evidence indicates such a crossover in studies of the 
magnetization\cite{Schilling93}
and muon-spin rotation.\cite{Aegerter96}  Furthermore, the unusual flux dynamics
have been shown to lead to an anomalously low irreversibility 
field, $H_{irr}$,\cite{huse92} below which 
resistivity drops in the superconductor. \cite{Schilling93,malozem88}
The value of $H_{irr}$ forms an irreversibility line in the H-T plane, 
which is a key feature for understanding flux-line dynamics.  Much
theoretical effort has been expended to tie this feature to the physical 
properties 
of the high-temperature superconductors.
More recently, experiments by Fuchs and coworkers \cite{fuchs98a,fuchs98b} have 
clarified the
 situation by demonstrating that, in 
Bi$_{2}$Sr$_{2}$CaCu$_{2}$O$_{8+\delta}$,  
$H_{irr}$ is determined by surface 
barriers.  We capitalize 
on this experimental fact to demonstrate the role played by interplanar spacing 
on the formation of flux lines.

Previous investigations of the role of anisotropy have shown shifts of the 
irreversibility and melting lines
 in oxygen-reduced Bi$_{2}$Sr$_{2}$CaCu$_{2}$O$_{8+\delta}$.
\cite{khaykovich96,Kishio94,mumtaz98,ooi98}  
Oxygen annealing simultaneously produces these four physical changes in the 
sample: (1) the {\it c} -axis lattice spacing, 
(2) $T_c$, (3) the in-plane penetration depth, $\lambda_{ab}$, and (4) the 
density of pinning sites.
A typical annealing study achieves a reduction in the {\it c} -axis lattice 
parameter of  
roughly 8 pm, a 0.3\% change,
at the cost of altering $T_c$ by 20\% or more.\cite{Kishio94,zhao98}  
Not unrelated, is the fact that ${\lambda}_{ab}$ at zero Kelvin has 
been shown to vary with oxygen doping, from 
210 nm to 305 nm.\cite{zhao98} 
At low temperatures, the situation is further complicated by the influence of 
bulk pinning.  Thus, in a doping study, 
the effects of interplanar separation, penetration depth, and
pinning site density on the flux dynamics are all intermingled.  This problem is
partially addressed in a study by 
Tamegai {\it et al.}\cite{tamegai}, which reports shifts in the melting line 
with the application of pressure.  To better 
understand the irreversible flux motion, it is necessary to deconvolve 
these phenomena.  

In this letter, we present the results of a study in which we directly 
investigate the effect of varying the interplanar 
spacing on the irreversibility line.  This is shown to increase the interlayer 
coupling, but to negligibly change the 
intraplanar superconductivity.  In our study, the application of pressures
 up to 2.5 GPa decreases the {\it c} -axis by 50 pm (a factor of 
3 greater than the change in either the {\it a} or {\it b} -axis).  $H_{irr}$ is 
increased by a factor of 10 at 
high temperatures; $T_c$ is only changed by 4\%; and ${\lambda_{ab}}$(T) is only 
marginally altered.  
As a result, we are able to show clear evidence of a 3-D to 2-D 
crossover in the flux dynamics and demonstrate a
significant pressure-induced change in the interplanar coupling.  

The Bi$_{2}$Sr$_{2}$CaCu$_{2}$O$_{8+\delta}$ single crystal used in this study
was grown by a self-flux technique using a stoichiometric ratio 
(Bi:Sr:Ca:Cu=2:2:1:2) of cations. \cite{y-doped,kend96} 
The crystal shape is that of a platelet, with dimensions 200 $\times$ 200 
$\times$ 50 $\mu$m$^3$ and a $T_c$ of
86.3 K.  Quasi-hydrostatic pressure is applied to the sample using a diamond 
anvil cell with a 
4:1 methyl-ethyl alcohol solution as the pressure-transmitting medium.  The 
pressure is applied and measured 
at room temperature and a calibration is used to determine the pressure at low 
temperatures to within
an uncertainty of $\pm0.3$ GPa.  

The irreversible flux motion is detected by measuring the third harmonic
of the {\it ac } susceptibility with primary and 
secondary coils wound around the diamond facets.  Both the 
{\it ac}- and {\it dc}-magnetic fields are applied parallel to the {\it c} -
axis, which is also parallel 
to the cylinder axis of the pressure cell. 
The {\it ac}-field amplitude is 0.5 mT, and the excitation
frequency is 3.7 kHz.  Details of this technique and of the diamond anvil cell 
are given in references
\cite{raphael98} and \cite{kim94}.

The nonlinear response to irreversible flux motion in the superconductor is 
shown in
Figure 1. 
The irreversibility line is defined by the locus of points determined 
by H and the onset temperature T$_1$ (see Figure 2).  In the past 
several years, a great deal of
progress has been made toward understanding the physical origins of the 
irreversibility line.  Often, this line does 
not indicate a phase boundary, but is simply a dividing line between reversible 
and irreversible flux motion, 
which is limited by extrinsic factors such as geometrical barriers, surface 
barriers, or pinning.  
At high temperatures, the irreversibility line has been shown to lie both above 
and below
the melting line, while extending well into the high field regime at 
low temperatures.\cite{Schilling93,majer95}  Furthermore, the onset of 
irreversibility has been shown to be determined by the barrier energy for flux 
entry into the superconductor.  
\begin{figure}
\epsfclipon
\centerline{
\epsfxsize= 3.2 in
\epsfbox{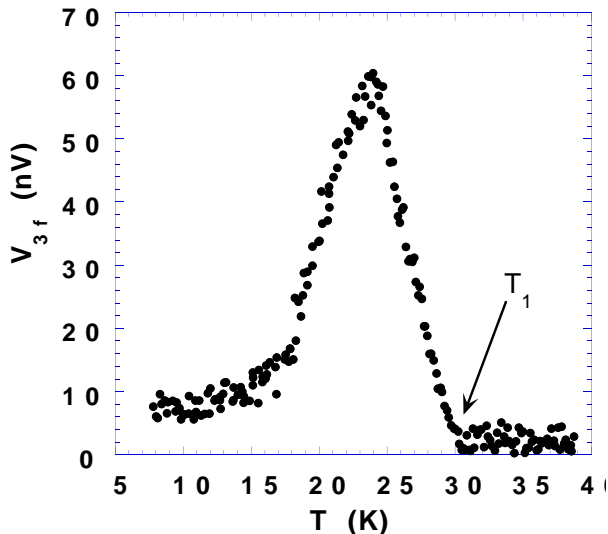}}
\caption{The third harmonic peak in Bi$_{2}$Sr$_{2}$CaCu$_{2}$O$_{8}$ when there 
is no applied pressure and the applied field is 3.0 Tesla.}  
\label{Figure 1}
\epsfclipoff
\end{figure}
The data of Figure 2 show a crossover in the pressure and temperature dependence 
of $H_{irr}$ near 50 mT.
These results are consistent with the muon-spin-resonance ($\mu$sr) 
data of Aegerter {\it et al.}\cite{Aegerter96} 
who find a 3-D to 2-D crossover field close to 70 mT.  These authors further 
show that the crossover field 
is reduced to 30 mT by an increase in $\lambda_{ab}$ and is independent of the 
anisotropy.  
At low temperatures, two-dimensional behavior is expected to occur, and 
$H_{irr}$ 
shows a weak pressure dependence.  Above the crossover temperature 
near 60 K,
the application of pressure significantly shifts the irreversibility line.  

We first focus on the low-temperature 
regime, where the data are described well by the theoretical model of Burlachkov 
{\it et al.}\cite{Burlachkov94} Here, 
the essential assumption is that the irreversible behavior
is a result of vortex pancakes penetrating surface barriers.  For high fields, 
much larger than the first penetration field 
(H $\gg$ H$_P$  $\approx$ 15 mT) and $T>T_o$ (defined below), 
the irreversibility field assumes an exponential form,  
\begin{equation}
H_{irr} \approx H_{c2}(T_o/2T)exp(-2T/T_o), 
\end{equation}
\begin{equation}
T_o =  {\phi_0^2 d \over (4 \pi \lambda_{ab})^2 ln(t/t_o)},
\end{equation}
$H_{c2}$ is the upper critical field, $\phi_0$ is the fluxon, {\it d }is the 
interlayer spacing, and 
$t$ and $t_o$ are time scales related to the rate of flux creep over the surface 
barrier. \cite{Kopylov90}
Here we equate the fractional change in the interlayer spacing with that of the 
{\it c} -axis 
obtained from compressibility data.\cite{Olsen91}
Then, we are able to determine $T_{o}$ by fitting our data to Eq. (1) as shown 
in Figure 2.
  For 0 GPa, 1.5 GPa, and 2.5 GPa, we obtain for $T_o$ values of 20.6 K, 23.5 K, 
and 
22.9 K, respectively ($\pm$ 2 K).  A constant value of $H_{c2} \approx 180 T$ is 
used here, and we obtain 
similar values of $T_{o}$ over a range of
reasonable, constant values for $H_{c2}$  ($50 T<H_{c2}<250 T$).    The 0 GPa 
and the 1.5 GPa data are indistinguishable 
while the irreversibility line at 2.5 GPa
is shifted to slightly higher temperatures.  Also note that the measured range 
of $T_{o}$ values corresponds to
a variation in $\lambda_{ab}$(T) of only 15 nm.  This indicates that the 
pressure has little effect
on the penetration depth. (Here we have taken $ln(t/t_o)$ to be 30 as in Ref. 
\cite{Burlachkov94}.)
\begin{figure}
\centerline{
\epsfclipon
\epsfxsize=3.2 in
\epsfbox{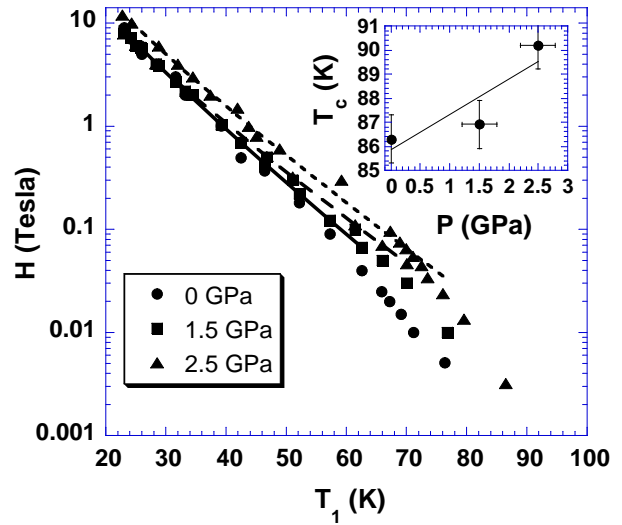}}
\caption{$H_{irr}$ at various pressures. At high fields the data 
show an exponential dependence which 
is expected for vortex pancakes penetrating the surface barrier.  
The inset shows $T_{c}$ as a function of pressure. The slope,
$dT_{c}/dP$ is consistent with that observed in other 
laboratories.\cite{klotz93} }
\label{Figure 2}
\epsfclipoff
\end{figure}
This result is illustrated in Figure 3 by a plot of $\lambda_{ab}$ 
{\it vs. c -}axis from pressure and 
from doping studies.  For the latter studies, $\lambda_{ab}$(0) is 
determined from magnetization measurements
\cite{zhao98,li96,Wald96}; for our study from fits to Eq. 2.  
The $\mu$sr data of Aegerter {\it et al.}\cite{Aegerter96} 
are not shown, but are consistent with the magnetization data of Li and 
coworkers.\cite{li96}  It is clearly shown in Figure 3 
that our experiment probes the effect of changing the interplanar spacing while 
holding the 
superconducting properties of the planes nearly constant.  
In contrast, oxygen-doping experiments probe the effect of 
modifying the intraplanar-superconducting order parameter, while causing 
relatively small 
changes in the interplanar spacing.
\begin{figure}
\centerline{
\epsfclipon
\epsfxsize=3.2 in
\epsfbox{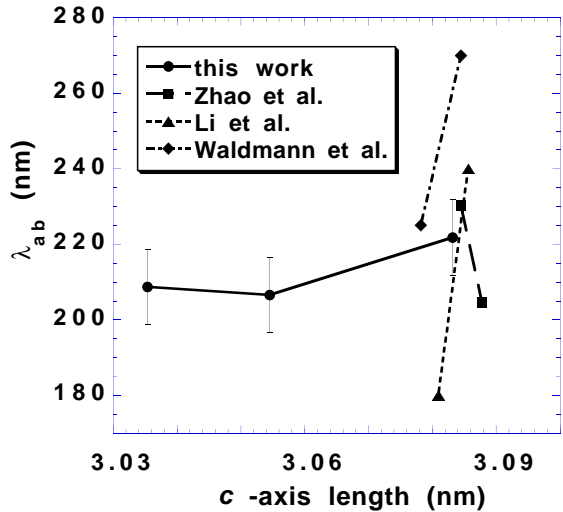}}
\epsfclipoff
\caption{Comparison of the change in $\lambda_{ab}$ with {\it c - }axis for 
pressure-induced changes [circles] and for oxygen-doping-induced changes: 
squares\cite{zhao98}, triangles\cite{li96}, 
diamonds\cite{Wald96}.  For our data, the {\it c - }axis 
is calculated from the pressure and the elastic moduli\cite{Olsen91}, 
and for the other data, from correlations between $T_{c}$, oxygen 
deficiency, and the {\it c - }axis spacing.\cite{Sun97}}
\label{Figure 3}
\end{figure}
Thus far, we have only discussed surface-barrier penetration as the mechanism to 
explain 
the irreversibility line.  There are other models that we have considered, 
which are rooted in bulk properties 
and which predict a power-law dependence for the irreversibility line: 
\begin{equation}
H_{irr} = H_0(1 - (T/T_c)^n)^\alpha.
\end{equation}
The above result holds for the irreversibility line denoting a flux-lattice-
melting 
transition (n = 1, $\alpha \leq$ 2) \cite{khaykovich96},  a 
Bose-glass transition (n = 1, $\alpha$ = 2 or 4/3) \cite{kim95,Xu91}, 
or a bulk-interplanar-decoupling transition of the vortices 
(n = -1, $\alpha$ = 1) \cite{khaykovich96}.  Our data can be represented 
by these models only for large values 
of the exponent ($\alpha$ = 7.4 for T $<$ 60 K and $\alpha$ = 3.5 for T $>$ 70 
K) or 
for unphysically large values of the scaling fields H$_0$.  
This is similar to results obtained by Schilling and 
coworkers.\cite{Schilling93}  
Thus we conclude that the irreversibility-line data are not indicative of a bulk 
transition in the sample.

In contrast to the low-temperature data, our high-temperature data 
show a significant pressure effect.  
Moreover, the closer these data are to $T_c$, the steeper 
the temperature dependence.  In a 
strictly 2-D interpretation, this would 
indicate a decrease in the penetration depth for T $>$ 60K (see Eq. 2): 
a physical impossibility.  Thus, we are led to 
conclude that 
this stiffening of the irreversibility line is due to the onset of 3-D 
coupling between the vortices.  
Both the change in the temperature dependence and 
the stronger pressure effect reinforce this interpretation.  
\begin{figure}
\centerline{
\epsfclipon
\epsfxsize=3.2 in
\epsfbox{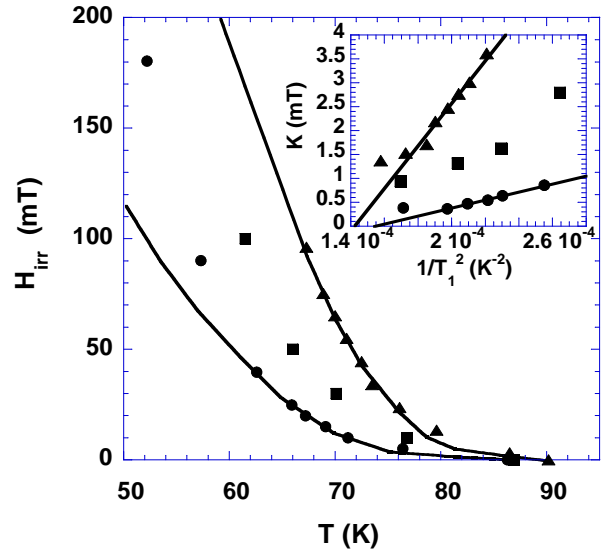}}
\caption{Fit to the high-temperature data for pressures of 0 GPa 
(circles), 1.5 GPa (squares) and 2.5 GPa (triangles).  The linear dependence 
seen 
in the inset is that expected for individual vortex lines penetrating
the surface barrier, and the deviation from this fit at high temperatures is 
observed 
for applied fields close to $H_{c1}$.} 
\label{Figure 4}
\end{figure}
One would then be led to analyze the data in terms of a model based on 
the penetration of the surface barrier by individual 3-D fluxoids.  For this 
case, the 
results of Burlachkov {\it et al.} indicate that the 
irreversibility line is described by the following expression 
\cite{Burlachkov94}:
\begin{equation}
{H_{irr} \over{Z^2(T) ln^{3}(H_{c2}/H_{irr})}} \approx {\pi \over 256 \gamma}  
{\phi_0  T_{o}^2 \over d^2} {1 \over T^2}  
\end{equation}
where $\gamma=(m_{c}/ m_{ab})^{1/2} = 
\lambda_{c}/\lambda_{ab}$ is the effective-mass-anisotropy
 parameter, and $H_{c2}(T)$ is linear with a slope of -2.7 T/K \cite{Li93}.  
$Z(T) = \lambda_{ab}^2(0)/\lambda_{ab}^2(T)$  is the temperature 
dependence of the penetration depth, taken from the
data of Waldmann {\it et al.} \cite{Wald96} 
  In Figure 4 we show the data and fits, and in the inset we 
linearize the data by plotting the left-hand side of Eq. 4 (K in the Figure) 
{\it vs.} 1/T$^2$.  At 0 GPa the data
show a linear dependence for $10<H<40$ mT and $62.6<T_{1}<71$ K.  
Not enough data were measured at
1.5 GPa to justify a fit, but the increase in slope is apparent. 
At 2.5 GPa the slope continues to increase, with
 the data showing a linear dependence for $24<H<96$ mT and 
 $67<T_{1}<76$ K.  As demonstrated by
the low-temperature data, the pressure does not significantly alter 
$\lambda_{ab}$.  
Therefore, the increase in slope
is due to a pressure-induced decrease in the effective-mass anisotropy of a 
factor of four.  The value of the anisotropy is difficult to measure in 
Bi$_{2}$Sr$_{2}$CaCu$_{2}$O$_{8+\delta}$, because of its two-dimensional 
electronic properties.  Using the fitted values of $T_{o}$ from above, 
gives $\gamma$ values of 530$\pm$100 and 130$\pm$20 for 
0 and 2.5 GPa, respectively.  At 0 GPa, this is significantly higher 
than other reports in the literature, except for a 
value of 370 measured by Schilling and coworkers \cite{Schilling93} 
in a similar field range.  Our result is further supported by the 
observation of $\gamma$ increasing rapidly with decreasing field \cite{kaw98} 
and the lower limit of  $\gamma$ $>$ 150 from the torque 
magnetometry measurements of Mart\'{i}nez and coworkers \cite{mart92}. 

In a related work, Tamegai {\it et al.} \cite{tamegai} report a pressure-dependent 
stiffening of the vortex-lattice melting line up to a pressure of 1 GPa.  
Their results indicate that the melting field shifts at a rate, 
$\Delta B_{m}(P)/B_{m}(0)$, of $33\%/$GPa at 65 K.
For $B_{m} \propto 1/\gamma^{2}$, this corresponds to a $26\%$ decrease in 
in $\gamma$ at 2.5 GPa, which is smaller than what we obtain using Eq. 4.

Nevertheless, our results clearly show that the interplanar coupling 
of the vortices changes dramatically with the application of pressure.  
Such a drastic increase in the interplanar coupling has also been measured 
by Yurgens {\it et al.}, who observe a large pressure
dependence of the {\it c} -axis critical current, $I_{c}$, 
in Bi$_{2}$Sr$_{2}$CaCu$_{2}$O$_{8+\delta}$.\cite{yurgens}
At $\sim$65K and $H=0$ mT, the authors report a relative change in $I_{c}$, 
$\Delta I_{c}(P)/I_{c}(0)$, of $\sim133\%/$GPa
for $\Delta P= 0.8$GPa.  Their experiment directly probes the interplanar 
Josephson effect in
Bi$_{2}$Sr$_{2}$CaCu$_{2}$O$_{8+\delta}$, while in our experiment the role of
Josephson coupling is manifested through an effective increase in the height of 
the surface barrier.

In total, the results that we have presented show that even modest changes in 
the {\it c} -axis lead to
dramatic effects on flux-line formation.  We observe that the applied pressure 
seems to have little 
influence on the superconducting
order parameter (as evidenced by the insensitivity of $\lambda_{ab}$ and of 
$T_c$ to pressure).  
By contrast, the application of pressure decreases the anisotropy  and increases 
the energy 
needed to bend an individual vortex line.  Thus, we demonstrate the importance 
of interplanar spacing on the 
formation of flux lines.  

Our experiment probes the superconducting 
properties in a very different manner than is done in doping studies.  In the 
pressure 
experiments (up to our maximum pressure 
of 2.5 GPa), the intraplanar superconductivity seems to be relatively unchanged, 
while the coupling between planes is strongly 
affected.  This contrasts to doping experiments where the major effect seems to 
be to alter the 
superconducting order parameter, 
while causing only modest changes in interplanar spacing.  Doping does effect 
the anisotropy, but mainly 
by changing the magnetic penetration depth.

In a general way, this experiment sheds light on the role of surface barriers to 
flux penetration in 
determining the position 
of the irreversibility line.  The consistency of the temperature and pressure 
dependencies of $H_{irr}$ shows clear 
evidence that there are two regimes of flux motion.  For temperatures
below about 60K, the flux configuration is that of two-dimensional pancake 
vortices.  
This crosses over to one of highly-anisotropic, 
three-dimensional flux tubes at higher temperatures.  The irreversibility line 
is then determined by the energy needed to push pancake 
vortices into the sample at low temperatures or to push line vortices into 
the sample at temperatures closer to $T_c$.  

This work is sponsored by the Office of Naval Research and the Defense Advanced 
Projects Research Agency.  
We acknowledge helpful discussions with T. P. Devereaux, J. H. Claassen, M. S. 
Osofsky, and R. J. Soulen, Jr.

\end{multicols}

\end{document}